\documentclass[10pt]{iopart}
\usepackage{iopams}
\usepackage{bm}
\usepackage{amssymb}
\usepackage{cite}
\usepackage{pst-all}
\input epsf
\usepackage{epic}
\usepackage{graphicx}
\usepackage{pstricks}

\newtheorem{lemma}{Lemma}
\newtheorem{proposition}{Proposition}

\def\be{\begin{equation}}
\def\ee{\end{equation}}

\def\bx{\bm x}
\def\by{\bm y}
\def\bz{\bm z}
\def\bn{\bm n}
\def\bN{\bm N}
\def\bX{\bm X}

\def\sA{\scriptscriptstyle A}
\def\sB{\scriptscriptstyle B}
\def\sAB{\scriptscriptstyle AB}

\begin{document}

\title[New method for determining the light travel time...]{New method for determining the light travel time in static, spherically symmetric spacetimes. \\
Calculation of the terms of order $G^3$}

\author{Bernard Linet$^1$ and Pierre Teyssandier$^2$}

\address{$^1$ Laboratoire de Math\'ematiques et Physique Th\'eorique, CNRS/UMR 7350, \\
F\'ed\'eration Denis Poisson, Universit\'e Fran\c{c}ois Rabelais, F-37200 Tours, France}
\address{$^2$ SYRTE, CNRS/UMR 8630, UPMC, Observatoire de Paris, 61 avenue de \\
 l'Observatoire, F-75014 Paris, France}
\eads{\mailto{Bernard.Linet@lmpt.univ-tours.fr},  \mailto{Pierre.Teyssandier@obspm.fr}}


\begin{abstract}

A new iterative method for calculating the travel time of a photon as a function of the spatial positions of the emitter and the receiver in the field of a static, spherically symmetric body is presented. The components of the metric are assumed to be expressible in power series in $m/r$, with $m$ being half the Schwarzschild radius of the central body and $r$ a radial coordinate.  The procedure exclusively works for a light ray which may be described as a perturbation in powers of $G$ of a Minkowskian null geodesic, with $G$ being the Newtonian gravitational constant. It is shown that the expansion of the travel time of a photon along such a ray only involves elementary integrals whatever the order of approximation. An expansion of the impact parameter in power series of $G$ is also obtained. The method is applied to explicitly calculate the perturbation expansions of the light travel time and the impact parameter up to the third order. The full expressions yielding the terms of order $G^3$ are new. The expression of the travel time confirms the existence of a third-order enhanced term when the emitter and the receiver are in conjunction relative to the central body. This term is shown to be necessary for determining the post-Newtonian parameter $\gamma$ at a level of accuracy of $10^{-8}$ with light rays grazing the Sun.
\end{abstract}



\section{Introduction} \label{intro}

Determining the travel time of a photon as a function of the positions of the emitter and the receiver for a given time of emission (or reception) is a crucial problem in many tests of general relativity. Indeed, such a function, that is called a time transfer function, is relevant for modelling not only the experiments involving the measurement of a time delay or the comparison of distant clocks \cite{blanchet,linet}, but also the bending of light and the highly accurate astrometry\cite{leponcin1,leponcin2,teyssandier2}. 

The aim of this paper is to determine the time transfer function in the exterior space of a static, spherically symmetric body as an asymptotic expansion in powers of the Newtonian gravitational constant $G$. So we neglect the multipole structure, the rotation and the dynamical aspects occurring in the realistic models. This limitation is justified since it is currently accepted that modelling the post-linear regime of solar experiments which can be planned in the foreseeable future only requires to take into account the mass of the Sun. Moreover, the influence of the cosmological constant is neglected.

We concentrate on a class of metric theories of gravity in which it is possible to suppose that the photons are propagating in a region where the components of the metric are analytic expansions in powers of $m/r$, where the monopole term $m$ is half the Schwarzschild radius of the central body and $r$ an isotropic radial coordinate. The metric is thus characterized by an infinity of dimensionless constants generalizing the well-known post-Newtonian parameters $\beta$ and $\gamma$. In fact, we restrict our attention to the case where the photon follows a path that we call a quasi-Minkowskian light ray (see \cite{teyssandier2}), that is a null geodesic described as a perturbation in powers of $G$ of a Minkowskian null segment. The corresponding time transfer function is then represented by a generalized post-Minkowskian expansion in powers of $G$. For the sake of brevity, a term of order $G^n$ will be said to be of order $n$.

The first-order term in the expansion of the time transfer function is the well-known Shapiro time delay \cite{shapiro}, which can be obtained by different reasonings, with some of them involving only elementary calculations (see, e.g., \cite{will} or \cite{blanchet}). Nevertheless, calculating the higher-order terms is a more difficult problem. As far as we know, two distinct approaches are currently available.

{\it a)} The method based on integrating the null geodesic equations. This procedure has been applied up to the order $G^2$ within the framework of the parametrized post-post-Newtonian formalism in \cite{richter}, and then reconsidered in the case of the exterior Schwarzschild metric for a class of quasi-Galilean coordinate systems of interest in celestial mechanics \cite{brumberg1,brumberg2}. The problem has been recently revisited in\cite{klioner1} for a three-parameter family of static, spherically symmetric spacetimes in the context of the Gaia mission. However, the approach developed in \cite{brumberg1,brumberg2,klioner1} presents the drawback to be indirect, since the results are deduced from a solution which corresponds to a light ray emitted at infinity in a given direction. 

{\it b)} The methods which are natively adapted to the generic case where both the emitter and the receiver of the light rays are located at finite distances from the origin of the spatial coordinates. These methods are based either on an iterative determination of the Synge or bi-scalar world function (see \cite{john} for the Schwarzschild metric and \cite{leponcin1} for a more general and simpler approach), or on an iterative integration of the Hamilton-Jacobi (or eikonal) equation satisfied by the time transfer function (see \cite{teyssandier1}, and \cite{ashby} for a more recent treatment). The case $n=2$ has been successfully solved using the two variants. 

The calculations required by the procedures mentioned in {\it a)} and {\it b)} get quickly unwieldy as $n$ is increasing. For this reason, we present here a new method enabling a systematic determination of the time transfer function up to any given order $n$. This approach relies on the null geodesic equations written in spherical variables. The calculation of the time transfer function is elementary when the light ray is radial. Consequently, a specific method for determining the time transfer function is needed only in the case where the ray is non-radial. The starting point is the fact that for a non-radial ray which is not passing through a pericentre between its emission and its reception, the light travel time is given by an integral involving only one unknown quantity, namely the impact parameter of the ray. This expression transforms into an integro-differential equation when the property of the impact parameter to be a derivative of the time transfer function is taken into account (see \cite{teyssandier2}). We show that this equation can be solved by an iterative procedure and that each perturbation term involved in the time transfer function is expressible as a sum of elementary integrals which are easy to calculate with any symbolic computer program. A theorem of analyticity of the perturbation terms proved in the present paper allows to extend the results to the more general cases where the light ray is passing through a pericentre. Since the property of analyticity we use is inferred from the above-mentioned Hamilton-Jacobi equation, it may be pointed out that our procedure is an hybrid of the above-mentioned methods.  

In order to illustrate the convenience of the new method, we carry out the explicit calculation of the time transfer function up to the order $G^3$. The full expression of the third-order term that we obtain is new. This expression markedly improves the result previously found in \cite{ashby}, which only yields a partial expression describing the asymptotic form of the time transfer function when the emitter and the receiver tend to be in conjunction. Let us emphasize that our formula must not be confused with the expression of order $G^3$ previously obtained in \cite{sarmiento} and \cite{keeton}. Indeed, the expressions of the light travel time obtained in these works involve the radial coordinate of the pericentre of the ray without explicitly calculating this quantity as a function of the spatial positions of the emitter and the receiver. 

It may be argued that the most accurate projects for testing general relativity, like SAGAS \cite{wolf}, ODYSSEY \cite{christophe}, LATOR \cite{turyshev} or ASTROD \cite{braxmaier}, are generally considered as only requiring a knowledge of the propagation of light up to the second order (see, e.g., \cite{minazzoli} and references therein). However, the occurrence of a so-called `enhanced term', that is the possibility for an effect of order $n+1$ to be greater than some contributions of order $n$ must be faced, as it is pointed out in \cite{ashby}. The necessity to be cautious has been recently shown for the Gaia mission. This mission, indeed, is currently tackled within the linearized, post-Minkowskian regime \cite{kopeikin1,kopeikin2} or the usual post-Newtonian approximation\cite{klioner2,crosta}. Nevertheless, an apparent discrepancy between the standard approach and the numerical integration of the null geodesic equations has required an in-depth discussion of an enhanced term of order $G^2$ (see \cite{klioner1} and \cite{teyssandier2}). For this reason, it seems to us that exploring new systematic procedures enabling the calculation of the time transfer function at any order is fully justified. Applying our results to experimental projects like SAGAS largely confirms this analysis. Indeed, we prove that the third-order term in the expansion of the time transfer function gives rise to an enhanced contribution in a solar conjunction. We show that this contribution must be taken into account in the attempts to determine the parameter $\gamma$ with an accuracy of $10^{-8}$.
 
The paper is organized as follows. Section \ref{notations} lists the notations and conventions we use. In section \ref{TTFsss}, the general assumptions on the metric are stated. Section \ref{TTFqM} is devoted to the expansion in a series in powers of $G$ of the time transfer function associated with a quasi-Minkowskian light ray. Section \ref{recurTT} yields a recurrence relation satisfied by the perturbations terms involved in the expansion of the time transfer function. A fundamental property of analyticity is established for these terms in section \ref{analytTT}. The impact parameter of any non-radial quasi-Minkowskian light ray is shown to be expressible as a series in powers of $G$ in section \ref{const_integr}. This feature is the basis of the new iterative procedure proposed in this paper for determining the time transfer function at any order $n$. This procedure is implemented in section \ref{princmeth}. A simplification is carried out in section \ref{pract_impl}. The perturbation expansion of the time transfer function is explicitly calculated up to the third-order in $G$ in section \ref{TTF3rd}. The appearance of enhanced terms at each order is shown in section \ref{Suit3rd}. The relevance of these terms is discussed for some solar system projects in section \ref{Apexp}. The impact parameter of the corresponding light ray is obtained as a function of the positions of the emitter and the receiver up to the order $G^3$ in section \ref{imp_par3}. The impact parameter of a ray emitted at infinity in an arbitrary direction and observed at a given point is derived in section \ref{rinfin}. Concluding remarks are given in section \ref{concl}. An appendix yields some hints for the hand calculation of the third-order term in the time transfer function.

\section{General assumptions, notations and conventions} \label{notations}

Our general assumptions, notations and conventions are the following.
\begin{itemize}

\item
Spacetime is assumed to be a static, spherically symmetric manifold $({\cal V}_4,g)$. We suppose that there exists a region ${\cal D}_h$ in which the metric $g$ is regular, asymptotically flat and may be interpreted as the gravitational field of a central body having a mass $M$. We put $m=GM/c^2$.

\item
 We assume that ${\cal D}_h$ may be entirely covered by a single quasi-Cartesian system of coordinates $(x^0,x^i)$ adapted to the symmetries of the metric. We put $x^0=ct$, with $t$ being a time coordinate, and ${\bm x}=(x^i)$.

\item
Greek indices run from 0 to 3, and latin indices run from 1 to 3.

\item 
The signature adopted for the metric is $+ - - -$.

\item
Any bold italic letter refers to an (ordered) triple: $(a^1,a^2,a^3)=(a^i)=\bm a$. All the triples are regarded as 3-vectors of the ordinary Euclidean space.

\item
Given two triples $\bm a$ and $\bm b$, $\bm a . \bm b$ denotes the Euclidean scalar product $a^i b^i$, with Einstein's convention on repeated indices being used. 

\item
$\vert \bm a \vert$ denotes the formal Euclidean norm of the triple $\bm a$: $\vert \bm a \vert=(\bm a . \bm a)^{1/2}$. If $\vert \bm a \vert=1$, $\bm a$ is conventionally called a unit (Euclidean) 3-vector.

\item
$\bm a\times\bm b$ is the triple obtained by the usual rule giving the exterior product of two vectors of the Euclidean space.

\item
Given a bi-scalar function $F(\bx, \by)$, $\bm \nabla_{\bx}F(\bx, \by)$ and  $\bm \nabla_{\by}F(\bx, \by)$ denote the gradients of $F$ with respect to $\bx$ and $\by$, respectively.

\end{itemize}

\section{Generalized post-Minkowskian expansion of the metric} \label{TTFsss}

For convenience, the coordinates $(x^0,{\bm x})$ are chosen so that the metric takes an isotropic form in the domain of regularity ${\cal D}_h$: 
\be \label{ds2}
ds^2 = {\cal A}(r)(dx^0)^2-{\cal B}^{-1}(r)\delta_{ij}dx^idx^j,
\ee
where $r=\vert\bx\vert$. Using the corresponding spherical coordinates $(r,\vartheta ,\varphi )$, one has 
$$
\delta_{ij}dx^idx^j = dr^2+r^2d\vartheta^2+r^2\sin^2\vartheta d\varphi^2.
$$

The light rays of the metric (\ref{ds2}) are also the light rays of any metric $d\tilde{s}^2$ conformal to (\ref{ds2}). This feature enables us to carry out our calculations for a metric containing only one potential. We choose $d\tilde{s}^2={\cal A}^{-1}(r)ds^2$, that is
\be \label{cds2}
d\tilde{s}^2=(dx^0)^2-{\cal U}(r)\delta_{ij}dx^idx^j,
\ee
where ${\cal U}$ is defined by
\be \label{Ur}
{\cal U}(r)=\frac{1}{{\cal A}(r){\cal B}(r)}.
\ee

The metric (\ref{ds2}) is considered as a generalization of the exterior Schwarzschild metric, which may be written in the form
\be \label{sm}
ds^2=\frac{\left(1-\displaystyle \frac{m}{2r} \right)^2}{\left(1+\displaystyle \frac{m}{2r}\right)^2} (dx^0)^2-
\left( 1+\frac{m}{2r}\right)^4\delta_{ij}dx^idx^j
\ee
in the region outside the event horizon located at $r=m/2$. So we henceforth assume that there exists a value $r_{h}>0$ of the radial coordinate such that the domain of regularity ${\cal D}_h$ is the region outside the sphere of radius $r_h$.  If there exists at least one event horizon, we must take for $r_{h}$ the value of $r$ on the outer horizon. By analogy with general relativity we consider that $r_h\sim m$ and we suppose that whatever $r>r_h$, ${\cal A}(r)$ and ${\cal B}^{-1}(r)$ are positive functions represented by analytical expansions as follow:
\numparts 
\begin{eqnarray} 
\fl {\cal A}(r) = 1-\frac{2m}{r}+2\beta \frac{m^2}{r^2}-\frac{3}{2}\beta_3\frac{m^3}{r^3}+\beta_4\frac{m^4}{r^4}+
\sum_{n=5}^{\infty} \frac{(-1)^nn}{2^{n-2}}\beta_n \frac{m^n}{r^n} , \label{ppnA}\\
\fl {\cal B}^{-1}(r) = 1+2\gamma\frac{m}{r}+\frac{3}{2}\epsilon \frac{m^2}{r^2}+\frac{1}{2}\gamma_3 \frac{m^3}{r^3}+
\frac{1}{16}\gamma_4 \frac{m^4}{r^4}+\sum_{n=5}^{\infty}(\gamma_n -1)\frac{m^n}{r^n} , \label{ppnB}
\end{eqnarray} 
\endnumparts 
where the coefficients $\beta,\beta_3,\dots ,\beta_n,\gamma ,\epsilon ,\gamma_3,\dots ,\gamma_n,\dots$ are generalized post-Newtonian parameters chosen so that
\be \label{GRbc}
\beta =\gamma =\epsilon =1, \qquad \beta_n=\gamma_n=1 \quad \mbox{for} \quad n\geq3
\ee
in general relativity. It results from (\ref{Ur}), (\ref{ppnA}) and (\ref{ppnB}) that the potential ${\cal U}(r)$ occurring in (\ref{cds2}) may be written as 
\be \label{invAB}
{\cal U}(r)= 1+2(1+\gamma)\frac{m}{r}+\sum_{n=2}^{\infty} 2\kappa_{n}\frac{m^n}{r^n},
\ee
for $r>r_h$, with the coefficients $\kappa_{n}$ being constants which can be expressed in terms of the generalized post-Newtonian parameters involved in the expansions of ${\cal A}(r)$ and ${\cal B}(r)$. Taking into account a notation already introduced in \cite{teyssandier2}, namely
\be \label{kappa}
\kappa=2(1+\gamma)-\beta+\mbox{$\frac{3}{4}$}\epsilon,
\ee
$\kappa_2$ and $\kappa_3$ are given by
\be \label{kappa3}
\kappa_2= \kappa,\quad
\kappa_3=2\kappa-2\beta(1+\gamma)+\mbox{$\frac{3}{4}$}\beta_3+\mbox{$\frac{1}{4}$}\gamma_3.   
\ee
In general relativity, we have
\be \label{GRk}
\kappa_2=\kappa=\mbox{$\frac{15}{4}$}, \quad \kappa_3=\mbox{$\frac92$}.
\ee

\section{Time transfer function for a quasi-Minkowskian light ray} \label{TTFqM}

Let $\bx_{\sA}$ and $\bx_{\sB}$ be two points located in ${\cal D}_{h}$.  Throughout this paper, we generically consider a photon emitted at $\bx_{\sA}$ and received at $\bx_{\sB}$. The time of emission and the time of reception of this photon are denoted by $t_{\sA}$ and $t_{\sB}$, respectively. It is assumed that the light ray followed by the photon is a null geodesic path $\Gamma$ which is entirely lying in the domain of regularity ${\cal D}_{h}$. As it has been emphasized in introduction, it would be of primordial interest for modelling a lot of relativistic effects to determine the time transfer function associated with $\Gamma$, that is, the expression giving the travel time $t_{\sB}-t_{\sA}$ as a function of $\bx_{\sA}$ and $\bx_{\sB}$: 
\begin{equation} \label{TABG}
t_{\sB}-t_{\sA}={\cal T}_{\Gamma}(\bx_{\sA},\bx_{\sB}).
\end{equation}

In practice, however, the problem is extremely complicated since there exists in general an infinite set of light rays emitted at  $\bx_{\sA}$ at a given time $t_{\sA}$ and passing through $\bx_{\sB}$ (see, e.g., \cite{darwin, luminet} for the Schwarzschild metric and \cite{giannoni} for a larger class of spacetimes). So, in this paper, we restrict our attention to the special class of null geodesic paths we have called the quasi-Minkowskian light rays in \cite{teyssandier2}. This means that in what follows, the path covered by the photon is assumed to be entirely confined in ${\cal D}_{h}$ and to be described by parametric equations of the form 
\begin{eqnarray}  
x^0=ct_{\scriptscriptstyle A}+\xi\vert\bm x_{\scriptscriptstyle B}-\bm x_{\scriptscriptstyle A}\vert+\sum_{n=1}^{\infty}X^0_{(n)}(\bm x_{\scriptscriptstyle A}, \bm x_{\scriptscriptstyle B}, \xi), \label{qM0} \\
\bx=\bz(\xi)+\sum_{n=1}^{\infty}\bm X_{(n)}(\bm x_{\scriptscriptstyle A}, \bm x_{\scriptscriptstyle B}, \xi),  \label{qMi}
\end{eqnarray}
where $\xi$ is the affine parameter varying on the range $0\leq \xi \leq 1$, $\bz(\xi)$ is defined by
\be \label{zxi}
\bz(\xi)=\bx_{\scriptscriptstyle A}+\xi (\bx_{\scriptscriptstyle B}-\bx_{\scriptscriptstyle A})
\ee
and the functions $X^0_{(n)}$ and $\bm X_{(n)}$ are terms of order $n$ obeying the boundary conditions 
\begin{eqnarray} 
X^0_{(n)}(\bm x_{\scriptscriptstyle A}, \bm x_{\scriptscriptstyle B}, 0)=0, \label{bcd0} \\ 
\bm X_{(n)}(\bm x_{\scriptscriptstyle A}, \bm x_{\scriptscriptstyle B}, 0)=\bm X_{(n)}(\bm x_{\scriptscriptstyle A}, \bm x_{\scriptscriptstyle B}, 1)=0. \label{bcd}
\end{eqnarray}
According to a notation already introduced in \cite{teyssandier2}, such a null geodesic path will be denoted by $\Gamma_{s}(\bx_{\sA},\bx_{\sB})$\footnote{In a static spacetime, the mention of the initial time $t_{\sA}$ may be omitted.}. For the sake of brevity, the time transfer function associated with $\Gamma_{s}(\bx_{\sA},\bx_{\sB})$ will be henceforth denoted by ${\cal T}(\bx_{\sA},\bx_{\sB})$ or simply by ${\cal T}$. Setting $\xi=1$ in (\ref{qM0}), it may be seen that this function can be expanded in power series of $G$ as follows:
\be \label{expT}
{\cal T}(\bx_{\sA},\bx_{\sB})=\frac{\vert \bx_{\sB}-\bx_{\sA} \vert}{c}+\sum_{n=1}^{\infty}{\cal T}^{(n)}(\bx_{\sA},\bx_{\sB}) ,
\ee
where ${\cal T}^{(n)}$ stands for the term of order $n$. 

Expansion (\ref{expT}) is easy to determine when $\bx_{\sA}$ and $\bx_{\sB}$ are linked by a radial null geodesic entirely lying in ${\cal D}_h$. In this case, indeed, it is immediately deduced from (\ref{cds2}) that the expression of ${\cal T}$ is given by the exact formula
\be \label{TABrd}
{\cal T}(r_{\sA},r_{\sB}) = \mbox{sgn$(r_{\sB}-r_{\sA})$}\frac{1}{c}\int_{r_{\sA}}^{r_{\sB}} \sqrt{{\cal U}(r)}dr,
\ee
where $r_{\sA}=\vert\bx_{\sA}\vert$ and $r_{\sB}=\vert\bx_{\sB}\vert$. Substituting for ${\cal U}(r)$ from (\ref{invAB}) into (\ref{TABrd}) shows that ${\cal T}$ may be expanded as follows:  
\be \label{expTr}
{\cal T}(r_{\sA},r_{\sB})=\frac{\vert r_{\sB}-r_{\sA}\vert}{c}+\sum_{n=1}^{\infty}{\cal T}^{(n)}(r_{\sA},r_{\sB}),
\ee
where the first three perturbation terms are given by
\begin{eqnarray} 
\fl {\cal T}^{(1)}(r_{\sA},r_{\sB}) =\frac{(1+\gamma)m}{c}\left\vert \ln\frac{r_{\sB}}{r_{\sA}}\right\vert, \label{Tr1} \\
\fl {\cal T}^{(2)}(r_{\sA},r_{\sB})=\left[\kappa-\mbox{$\frac{1}{2}$}(1+\gamma)^2\right]\frac{m^2}{r_{\sA} r_{\sB}}\frac{\vert r_{\sB}-r_{\sA}\vert}{c}, \label{Tr2} \\
\fl {\cal T}^{(3)}(r_{\sA},r_{\sB})=\mbox{$\frac{1}{2}$}\left[\kappa_3-(1+\gamma)\kappa+\mbox{$\frac{1}{2}$}(1+\gamma)^3\right]\frac{m^3}{r_{\sA} r_{\sB}}\left(\frac{1}{r_{\sA}}+\frac{1}{r_{\sB}}\right)\frac{\vert r_{\sB}-r_{\sA}\vert}{c}. \label{Tr3}
\end{eqnarray}  

Determining the right-hand side of (\ref{expT}) is much more complicated when $\Gamma_{s}(\bx_{\sA},\bx_{\sB})$ is not a radial geodesic. As it has been recalled in introduction, the perturbations terms ${\cal T}^{(n)}$ might be obtained by an iterative integration of the null geodesic equations. Indeed, taking into account that $d\tilde{s}^2=0$ along a null geodesic, it results from (\ref{cds2}) and (\ref{qMi}) that the time transfer function is given by  
\be \label{TABg}
\fl{\cal T}(\bx_{\sA},\bx_{\sB})=\frac{1}{c}\int_{0}^{1}\sqrt{{\cal U}(r(\xi))}\left\vert  \bx_{\sB}-\bx_{\sA}+\sum_{n=1}^{\infty}\frac{d\bX_{(n)}(\bm x_{\scriptscriptstyle A}, \bm x_{\scriptscriptstyle B}, \xi)}{d\xi}\right\vert d\xi,
\ee
where the integral is taken along $\Gamma_{s}(\bx_{\sA},\bx_{\sB})$. Taking into account the boundary conditions (\ref{bcd}), it may be inferred from (\ref{TABg}) that each function ${\cal T}^{(n)}$ is theoretically calculable if the perturbations terms $\bm X_{(1)},\dots, \bm X_{(n-1)}$ involved in (\ref{qMi}) are determined by solving the null geodesic equations. This procedure is cumbersome, however. So we develop a different method, based on a property of analyticity of the functions ${\cal T}^{(n)}$ which can be deduced from a recurrence relation (see sections \ref{recurTT} and \ref{analytTT}). 

\section{Recurrence relation satisfied by the functions ${\cal T}^{(n)}$} \label{recurTT}

It follows from a general result established in \cite{teyssandier1} that ${\cal T}(\bx_{\sA},\bx)$ satisfies a Hamilton-Jacobi equation which is equivalent to the eikonal equation
\be \label{HJ}
c^2 \vert\bm \nabla_{\bx}{\cal T}(\bx_{\sA},\bx)\vert^2 = {\cal U}(r)
\ee
when the metric is given by (\ref{cds2}). Replacing ${\cal T}(\bx_{\sA},\bx)$ in (\ref{HJ}) by its expansion in powers of $G$ and ${\cal U}(r)$ by (\ref{invAB}), and then applying the formulae already found in \cite{teyssandier1}, we get a proposition as follows.

\begin{proposition} \label{itTAB}
The perturbation terms ${\cal T}^{(n)}$ involved in expansion (\ref{expT}) may be written in the form
\be \label{TnFn}
{\cal T}^{(n)}(\bx_{\sA},\bx_{\sB})=\frac{1}{c}\vert \bx_{\sB} -\bx_{\sA} \vert F^{(n)}(\bx_{\sA},\bx_{\sB}),
\ee 
where the functions $F^{(n)}$ are determined by the recurrence relation 
\begin{eqnarray}  
\fl F^{(1)}(\bx_{\sA},\bx_{\sB})=(1+\gamma)m \int_{0}^{1}\frac{d\xi}{\vert \bz(\xi)\vert}, \label{FAB1} \\
\fl F^{(n)}(\bx_{\sA},\bx_{\sB})=\kappa_n m^n \int_{0}^{1}\frac{d\xi}{\vert \bz(\xi)\vert^n} \nonumber \\
-\frac{c^2}{2}\sum_{p=1}^{n-1}\int_{0}^{1}\left[\bm \nabla_{\bx}{\cal T}^{(p)}(\bx_{\sA},\bx).\bm \nabla_{\bx}{\cal T}^{(n-p)}(\bx_{\sA},\bx)\right]_{\bx=\bz(\xi)} d\xi \label{FABn}
\end{eqnarray}
for $n\geq2$, with $\bz(\xi)$ being defined by (\ref{zxi})\footnote{Let us emphasize that the integrals involved in (\ref{FAB1}) and (\ref{FABn}) are taken along the straight segment described by the parametric equation 
$\bx=\bz(\xi), 0\leq\xi\leq1$.}.
\end{proposition} 

The recurrence relation explicitly given in proposition \ref{itTAB} shows that $\Gamma_{s}(\bx_{\sA},\bx_{\sB})$ is unique provided that expansion (\ref{expT}) is an admissible representation of the time transfer function. However, determining the most general conditions under which our construction is valid remains an open problem. According to \cite{teyssandier1}, (\ref{FAB1}) and (\ref{FABn}) are inferred from an integro-differential equation involving the analytic expansion of the metric along the straight segment joining $\bx_{\sA}$ and $\bx_{\sB}$. Consequently, we shall henceforth assume that the expression $\vert \bx_{\sB}-\bx_{\sA}\vert [1+\sum_{p=1}^{n} F^{(p)}(\bx_{\sA},\bx_{\sB})]/c$ constitutes a reliable approximation of the time transfer function as long as the straight segment joining $\bx_{\sA}$ and $\bx_{\sB}$ does not intersect the hypersurface $r=r_h$, a condition expressed by the inequality
\be \label{strl}
\vert \bz(\xi) \vert >r_h \quad \mbox{for}\quad 0\leq \xi \leq 1.
\ee
This condition is largely satisfied for a star observed in the solar system, as well as in any foreseeable test of general relativity in the vicinity of the Sun.

It has been shown in \cite{teyssandier1} that a recurrent relation equivalent to the one stated by proposition \ref{itTAB} enables to carry out the calculation of ${\cal T}^{(1)}$ and ${\cal T}^{(2)}$. As matters stand, however, we do not know whether (\ref{FABn}) allows explicit computations for $n\geq3$. Going deeper into this question is beyond the scope of this paper. Proposition \ref{itTAB} is used here only for proving a property of analyticity which is indispensable for justifying our new procedure.  

\section{Analyticity of the functions ${\cal T}^{(n)}$} \label{analytTT}

Let us begin with proving the following lemma.

\begin{lemma} \label{analytF}
The functions $F^{(n)}$ recursively determined by (\ref{FAB1}) et (\ref{FABn}) are analytic in $\bx_{\sA}$ and $\bx_{\sB}$, except when $\bx_{\sA}$ and $\bx_{\sB}$ are such that $\bn_{\sB}=-\bn_{\sA}$, with $\bn_{\sA}$ and $\bn_{\sB}$ being defined as 
\be \label{nAnB}
\bn_{\sA}=\frac{\bx_{\sA}}{r_{\sA}}, \quad \bn_{\sB}=\frac{\bx_{\sB}}{r_{\sB}}. 
\ee.  
\end{lemma}

{\bf Proof of lemma \ref{analytF}.} The proposition is obviously true for $n=1$, since the integrand $1/\vert \bz(\xi)\vert$ in (\ref{FAB1}) is analytic in $\bx_{\sA}$ and $\bx_{\sB}$ for any $\xi$ such that $0\leq\xi\leq1$, provided that $\bn_{\sB}\neq -\bn_{\sA}$. Suppose now the validity of lemma \ref{analytF} for $F^{(1)},\dots,F^{(n)}$. Assuming $p$ to be such that $1\leq p\leq n$, and then substituting $\bz(\xi)$ for $\bx$ into $\bm \nabla_{\bx}{\cal T}^{(p)}(\bx_{\sA},\bx)$, it is immediately inferred from (\ref{TnFn}) that 
\begin{eqnarray}  
\fl c\left[\bm \nabla_{\bx}{\cal T}^{(p)}(\bx_{\sA},\bx)\right]_{\bx=\bz(\xi)}=\bN_{\!\sA\sB}F^{(p)}(\bx_{\sA},\bz(\xi))\nonumber \\
+\xi\vert\bx_{\scriptscriptstyle B}-\bx_{\scriptscriptstyle A}\vert\left[\bm \nabla_{\bx}F^{(p)}(\bx_{\sA},\bx)\right]_{\bx=\bz(\xi)}, \label{nabpT}
\end{eqnarray}
where $\bN_{\!\sA\sB}$ is defined by 
\be \label{NAB}
\bN_{\!\sA\sB}=\frac{\bx_{\sB}-\bx_{\sA}}{\vert\bx_{\sB}-\bx_{\sA}\vert} .
\ee
Using (\ref{nabpT}) leads to
\begin{eqnarray}  
\fl c^2\left[\bm \nabla_{\bx}{\cal T}^{(p)}(\bx_{\sA},\bx).\bm \nabla_{\bx}{\cal T}^{(n+1-p)}(\bx_{\sA},\bx)\right]_{\bx=\bz(\xi)}=F^{(p)}(\bx_{\sA},\bz(\xi))F^{(n+1-p)}(\bx_{\sA},\bz(\xi))\nonumber \\
+\xi (\bx_{\sB}-\bx_{\sA}).\left[F^{(p)}(\bx_{\sA},\bx)\bm \nabla_{\bx}F^{(n+1-p)}(\bx_{\sA},\bx)\right. \nonumber \\
+\left. F^{(n+1-p)}(\bx_{\sA},\bx)\bm \nabla_{\bx}F^{(p)}(\bx_{\sA},\bx)\right]_{\bx=\bz(\xi)} \nonumber \\
+\xi^2\vert\bx_{\sB}-\bx_{\sA}\vert^2\left[\bm \nabla_{\bx}F^{(p)}(\bx_{\sA},\bx).\bm \nabla_{\bx}F^{(n+1-p)}(\bx_{\sA},\bx)\right]_{\bx=\bz(\xi)}. \label{naTnaT}
\end{eqnarray}
It follows from our assumption that the right-hand side of (\ref{naTnaT}) is a sum of functions which are analytic in $\bx_{\sA}$ and $\bx_{\sB}$ for any $\xi$ such that $0\leq\xi\leq1$, except if $\bn_{\sB}=-\bn_{\sA}$. Each integral 
\[
\int_{0}^{1}\left[\bm \nabla_{\bx}{\cal T}^{(p)}(\bx_{\sA},\bx).\bm \nabla_{\bx}{\cal T}^{(n+1-p)}(\bx_{\sA},\bx)\right]_{\bx=\bz(\xi)}d\xi
\]
is therefore analytic if $\bn_{\sB}\neq -\bn_{\sA}$. The same property is obviously possessed by the integral $\int_{0}^{1}d\xi/\vert\bz(\xi)\vert^{n+1}$. Lemma \ref{analytF} is thus proved by recurrence.

Since $\vert\bx_{\scriptscriptstyle B}-\bx_{\scriptscriptstyle A}\vert$ is analytic except if $\bx_{\scriptscriptstyle B}\neq\bx_{\scriptscriptstyle A}$, we can state the proposition below.

\begin{proposition} \label{analytT}
The functions ${\cal T}^{(n)}
$ involved in expansion (\ref{expT}) are analytic in $\bx_{\sA}$ and $\bx_{\sB}$ when both the following conditions are met:

a) $\bx_{\sB}\neq\bx_{\sA}$;

b) $\bn_{\sB}\neq -\bn_{\sA}$.
\end{proposition}

The importance of this property will clearly appear in section \ref{princmeth}. It is worth of noting that the second condition in proposition \ref{analytT} is automatically fulfilled when inequality (\ref{strl}) is satisfied. This fact explains why the condition {\it b)} is never explicitly involved in the assumptions of the propositions enunciated below.

\section{Relation between ${\cal T}(\bx_{\sA},\bx_{\sB})$ and the impact parameter of $\Gamma_s (\bx_{\sA},\bx_{\sB})$}\label{const_integr}

\subsection{Null geodesic equations} 

Let $\Gamma$ be an arbitrary non-radial null geodesic path of the metric $d\tilde{s}^2$. We suppose that $\Gamma$ is confined in the region ${\cal D}_h$ and described by parametric equations $x^{\alpha}=x^{\alpha}(\zeta)$, where $\zeta$ is an arbitrarily chosen affine parameter. We choose the spherical coordinates $(r,\vartheta ,\varphi )$ so that $\vartheta = \pi /2$ for any point of this path. Denoting by $\tilde{l}_{\alpha}$ the covariant components of the vector tangent to $\Gamma_{s}(\bx_{\sA},\bx_{\sB})$, an equation as follows 
\be \label{ldx0}
\tilde{l}_0 dx^0+\tilde{l}_r dr+\tilde{l}_{\varphi}d\varphi=0
\ee
is satisfied along $\Gamma$ since $\tilde{l}_{\alpha}$ is a null vector. Owing to the symmetries of the metric, we have 
\begin{eqnarray} 
\tilde{l}_0=E, \label{ip1} \\
\tilde{l}_{\varphi}=-J, \label{ip2} 
\end{eqnarray}
with $E$ and $J$ being constants of the motion. For convenience, the affine parameter is chosen in such a way that $E>0$. Furthermore, it is always possible to suppose $J>0$ without lack of generality when calculating the time transfer function in a static, spherically symmetric spacetime. Then the quantity defined as
\be \label{db}
b=\frac{J}{E}
\ee
is the impact parameter of the light ray (see, e.g., \cite{chandra} and \cite{teyssandier2})\footnote{$b$ is an intrinsic quantity attached to $\Gamma$ since the constants of the motion $E$ and $J$ are themselves coordinate-independent quantities.}. It may be noted that $b=0$ would correspond to a radial null geodesic.

Since $d\tilde{s}^2=0$ along $\Gamma$, it follows from (\ref{ip1}), (\ref{ip2}) and (\ref{db}) that
\be \label{eqdf2}
\tilde{l}_r=-\varepsilon \frac{E}{r}\sqrt{r^2{\cal U}(r)-b^2},
\ee
where $\varepsilon=1$ when $r$ is an increasing function of time and $\varepsilon=-1$ when $r$ is a decreasing function of time\footnote{The sign of $\varepsilon$ in equation (\ref{eqdf2}) is changed if and only if the photon passes through a pericentre or an apocentre. The passage through an apocentre corresponds to an extreme relativistic case.}. Substituting for $\tilde{l}_r$ from (\ref{eqdf2}) into (\ref{ldx0}), and then dividing throughout by $E$, we obtain a relation enabling to determine the light travel time by an integration along 
$\Gamma (\bx_{\sA }, \bx_{\sB })$, namely
\be \label{dx0a}
dx^0 = b d\varphi+\frac{\varepsilon}{r}\sqrt{r^2{\cal U}(r)-b^2} dr.
\ee
 
Our procedure for calculating explicitly each function ${\cal T}^{(n)}$ rests on the property reminded in the next subsection that the impact parameter $b$ can be determined as a function of $\bx_{\sA}$ and $\bx_{\sB}$ by taking the partial derivative of ${\cal T}$ with respect to the cosine of the angle formed by $\bx_{\sA}$ and $\bx_{\sB}$.

\subsection{Expansion of the impact parameter as a series in powers of G} \label{select}

Let $[\varphi_{\sA}, \varphi_{\sB}]$ be the range of the angular function $\varphi (t)$ along a quasi-Minkowskian light ray $\Gamma_{s}(\bx_{\sA},\bx_{\sB})$. For the sake of brevity, we shall frequently use a notation as follows 

\be \label{mu}
\mu=\bn_{\sA}.\bn_{\sB}=\cos(\varphi_{\sB}-\varphi_{\sA}).
\ee
Using this notation, the time transfer function may be considered as a function of $r_{\sA}, r_{\sB}$ and $\mu$:
\[
{\cal T}(\bx_{\sA},\bx_{\sB})={\cal T}(r_{\sA},r_{\sB},\mu).
\]
It is then possible to enunciate the following proposition.

\begin{proposition} \label{bexpm}
Let $\bx_{\sA}$ and $\bx_{\sB}$ be two points in ${\cal D}_h$ such that both the conditions $\bn_{\sA} \neq \bn_{\sB}$ and (\ref{strl}) are fulfilled. The impact parameter $b$ of a quasi-Minkowskian light ray joining $\bx_{\sA}$ and $\bx_{\sB}$ may be expanded in powers of $G$ as follows:
\be \label{expb}
b=r_c\left[ 1+\sum_{n=1}^{\infty}\left( \frac{m}{r_c}\right)^n q_n\right],
\ee
where $r_c$ is the usual Euclidean distance between the origin $O$ of the spatial coordinates and the straight line passing through $\bx_{\sA}$ and $\bx_{\sB}$, namely
\be \label{rc}
r_c=\frac{r_{\sA}r_{\sB}}{\vert\bx_{\sB}-\bx_{\sA} \vert} \vert \bn_{\sA}\times \bn_{\sB} \vert,
\ee
and the quantities $q_n$ are functions of  $\bx_{\sA}$ and $\bx_{\sB}$ given by
\be \label{qn}
q_n=-c\left(\frac{r_{c}}{m}\right)^n\frac{\sqrt{1-\mu^2}}{r_c}\,\frac{\partial {\cal T}^{(n)}(r_{\sA},r_{\sB},\mu)}{\partial \mu}.
\ee
\end{proposition}


{\bf Proof of proposition \ref{bexpm}.}  Noting that
\[
\vert\bn_{\sA}\times\bn_{\sB}\vert = \sqrt{1-\mu^2},
\]
it is immediately inferred from equation (13) in \cite{teyssandier2} that the impact parameter of $\Gamma_{s}(\bx_{\sA}, \bx_{\sB})$ may be rewritten in the form
\be \label{bdT}
b=-c\sqrt{1-\mu^2} \, \frac{\partial {\cal T}(r_{\sA},r_{\sB},\mu)}{\partial \mu}.
\ee  
Substituting for ${\cal T}$ from (\ref{expT}) into (\ref{bdT}) directly leads to the expansion given by (\ref{expb}). The zeroth-order term is easily derived from the elementary formula
\be \label{R}
\vert \bx_{\sB}-\bx_{\sA} \vert =\sqrt{r_{\sA}^{2}-2r_{\sA}r_{\sB}\mu +r_{\sB}^{2}}. 
\ee
Indeed, using this expression and taking (\ref{rc}) into account yield 
\be \label{derR}
\frac{\partial \vert \bx_{\sB} - \bx_{\sA}\vert}{\partial  \mu}=-\frac{r_c}{\sqrt{1-\mu^2}}.
\ee

We shall see in the next section that the expression of the time transfer function corresponding to a quasi-Minkowskian light ray can be straightforwardly deduced from proposition \ref{bexpm}.

\section{Implementation of the method} \label{princmeth}

If $\Gamma_{s}(\bx_{\sA},\bx_{\sB})$ passes through a pericentre $\bx_{\scriptscriptstyle P}$, the integration of (\ref{dx0a}) requires the determination of the value of the radial variable at $\bx_{\scriptscriptstyle P}$ as a function of $\bx_{\sA}$ and $\bx_{\sB}$. The calculation of the time transfer function is very complicated for such a configuration. Fortunately, owing to the analytic extension theorem, it follows from proposition \ref{analytT} that it is sufficient to determine the expression of each term ${\cal T}^{(n)}$ as a function of $\bx_{\sA}$ and $\bx_{\sB}$ in an arbitrarily chosen open subset of the domain of analyticity. For this reason, the calculation of the functions ${\cal T}^{(n)}$ are henceforth carried out under the assumption that $\bx_{\sA}$ and $\bx_{\sB}$ fulfil the following conditions:

{\it a)} The radial variable $r$ along a quasi-Minkowskian null geodesic joining $\bx_{\sA}$ and $\bx_{\sB}$ is an increasing function of $t$:
\be \label{rinc}
\frac{dr}{dt} >0, \qquad t_{\sA}\leq t \leq t_{\sB}.
\ee

{\it b)} An inequality as follows
\be \label{cdc}
\bN_{\!\sAB} . \bn_{\sA} > 0
\ee
is satisfied, with $\bN_{\!\sAB}$ being defined by (\ref{NAB}).

These conditions considerably simplify the calculations. Indeed, (\ref{rinc}) eliminates the occurrence of any pericentre (or apocentre) between the emission and the reception of light and (\ref{cdc}) implies that the projection of the origin $O$ on the straight line passing through $\bx_{\sA}$ and $\bx_{\sB}$ lies outside the straight segment linking $\bx_{\sA}$ and $\bx_{\sB}$. One has therefore 
\be \label{rcrAB}
r_c <r_{\sA}\leq r \leq r_{\sB}
\ee
for any point of $\Gamma_{s}(\bx_{\sA},\bx_{\sB})$. These inequalities ensure that condition (\ref{strl}) is met, since $r_h<r_{\sA}$ for any point $\bx_{\sA}$ located in ${\cal D}_{h}$. 

Under these assumptions, integrating (\ref{dx0a}) along $\Gamma_{s}(\bx_{\sA},\bx_{\sB})$ is straightforward since the range of the angular function $\varphi(r)$ between the emission and the reception of the photon is given by
\be \label{phiAB}
\varphi_{\sB}-\varphi_{\sA}=\arccos\mu.
\ee
Noting that in this case $\varepsilon=1$, it may be seen that the time transfer function is then related to the impact parameter $b$ by an equation as follows
\be \label{tab}
{\cal T}(\bx_{\sA},\bx_{\sB})=\frac{1}{c}\left[b\arccos\mu+\int_{r_{\sA}}^{r_{\sB}}\frac{1}{r}{\sqrt{r^2{\cal U}(r)-b^2}}dr\right].
\ee

Since $b$ is a function of $\bx_{\sA}$ and $\bx_{\sB}$ determined by (\ref{bdT}), (\ref{tab}) has to be regarded as an integro-differential equation satisfied by ${\cal T}$. In order to solve this integro-differential equation by an iterative procedure, let us substitute (\ref{invAB}) for ${\cal U}$ and (\ref{expb}) for $b$. Expanding $\sqrt{r^2{\cal U}(r)-b^2}/r$ in a power series in $m/r_c$, rearranging the terms and introducing the notation
\be \label{s}
s=\sqrt{r^2-r_{c}^{2}},
\ee
we get an expression as follows for ${\cal T}$
\begin{eqnarray}
\fl{\cal T}(\bx_{\sA},\bx_{\sB})=\frac{1}{c}\left[r_c\arccos\mu+\int_{r_{\sA}}^{r_{\sB}}\frac{s}{r}\, dr\right] \nonumber \\
+\frac{1}{c}\sum_{n=1}^{\infty}\left(\frac{m}{r_c}\right)^n\Bigg\lbrace r_c q_n\arccos\mu
+\int_{r_A}^{r_B}\left[U_n-\frac{r_c^2q_n}{rs}\right]dr\Bigg\rbrace, \label{tABi}
\end{eqnarray}
where each $U_n$ is a function of $r$ which may be written in the form
\begin{eqnarray}  
U_1=\frac{(1+\gamma)r_c}{s}, \label{U1} \\
U_n=\sum_{k=0}^{3n-4}U_{kn}(q_1,\dots ,q_{n-1})r_c^{3n-k-2}
\frac{r^{k-n+1}}{s^{2n-1}} \label{Xn}
\end{eqnarray}
for $n\geq2$, with the quantities $U_{kn}(q_1,\dots ,q_{n-1})$ being polynomials in $q_1,\dots,q_{n-1}$. 
Noting that
\be \label{RAB1}
r_c\arccos\mu+\int_{r_{\sA}}^{r_{\sB}}\frac{s}{r} \, dr=\vert \bx_{\sB}-\bx_{\sA} \vert
\ee
and  
\be \label{fab0}
\arccos\mu -r_c\int_{r_{\sA}}^{r_{\sB}}\frac{dr}{rs}=0
\ee
when conditions (\ref{rinc}) and (\ref{cdc}) are met\footnote{Note that (\ref{RAB1}) is just (\ref{tab}) written in the case where the gravitational field vanishes, i.e., $m=0$.}, (\ref{expT}) is immediately recovered from (\ref{tABi}), with each perturbation term being given by
\be \label{TnAB}
{\cal T}^{(n)}(\bx_{\sA},\bx_{\sB})=\frac{1}{c}\left(\frac{m}{r_c}\right)^n\int_{r_A}^{r_B}U_n dr.
\ee

As it has been explained in the beginning of this section, the expression of ${\cal T}^{(n)}$ as a function of $\bx_{\sA}$ and $\bx_{\sB}$ derived from (\ref{TnAB}) can be regarded as valid even when conditions (\ref{rinc}) and (\ref{cdc}) are not met. In this sense, (\ref{TnAB}) constitutes the main ingredient of the procedure developed in this paper. 

The fact that the coefficient $q_n$ is not involved in $U_n$ and the property for each coefficient $q_k$ to be proportional to a derivative of the function ${\cal T}^{(k)}$ imply that ${\cal T}^{(n)}$ can be determined when the sequence of functions ${\cal T}^{(1)},\dots,{\cal T}^{(n-1)}$ is known. Moreover, it follows from (\ref{Xn}) that all the integrations involved in the right-hand side of (\ref{TnAB}) are elementary and can be carried out with any symbolic computer program. Consequently, our procedure enables us to perform the explicit calculation of ${\cal T}^{(n)}$ as a function of $\bx_{\sA}$ and $\bx_{\sB}$ whatever the order $n$.

The problem is treated here in a detailed manner up the third order. So, (\ref{U1}) must be supplemented by the expressions of $U_2$ and $U_3$, namely 
\begin{eqnarray} 
\fl U_2=-\frac{\kappa r_c^4}{rs^3}+\frac{(1+\gamma)q_1 r_c^3}{s^3}+\frac{[2\kappa-(1+\gamma)^2-q_1^2]r_c^2r}{2s^3},  \label{U2} \\
\fl U_3=\frac{\kappa_3 r_c^7}{r^2s^5} -\frac{\kappa q_1 r_c^6}{rs^5}- \frac{[2\kappa_3-(1+\gamma)(\kappa +q_1^2-q_2)]r_c^5}{s^5} \nonumber \\
+\frac{[2\kappa - 3(1+\gamma)^2-q_1^2+2q_2]q_1r_c^4r}{2s^5} \nonumber \\
+\frac{[2\kappa_3-(1+\gamma)(2\kappa - q_1^2-2q_2)+(1+\gamma)^3]r_c^3r^2}{2s^5}-\frac{q_1 q_2r_c^2 r^3}{s^5}.\label{U3}
\end{eqnarray} 

The expression of ${\cal T}^{(1)}$ is immediately inferred from (\ref{U1}) and (\ref{TnAB}). Noting that
\numparts 
\begin{eqnarray} 
\sqrt{r_{\sA}^{2}-r_{c}^{2}}=r_{\sA}\bN_{\sAB}.\bn_{\sA}, \label{NnA} \\
\sqrt{r_{\sB}^{2}-r_{c}^{2}}=r_{\sB}\bN_{\sAB}.\bn_{\sB} \label{NnB}
\end{eqnarray}
\endnumparts
when conditions (\ref{rinc}) and (\ref{cdc}) are met, we get
\be \label{Shap}
{\cal T}^{(1)}(\bx_{\sA}, \bx_{\sB})=\frac{(1+\gamma)m}{c}\ln\left(\frac{r_{\sB}+\bN_{\sAB}.\bx_{\sB}}{r_{\sA}+\bN_{\sAB}.\bx_{\sA}}\right).
\ee
As it could be expected, we recover the well-known Shapiro time delay expressed in isotropic coordinates (see, e.g., \cite{will}).

For determining $q_1$, it is preferable to rewrite (\ref{Shap}) in the more elegant form (see, e.g., \cite{blanchet})
\be \label{T1}
{\cal T}^{(1)}(\bx_{\sA}, \bx_{\sB})=\frac{(1+\gamma)m}{c}\ln \left(\frac{r_{\sA}+r_{\sB}+\vert \bx_{\sB} -\bx_{\sA} \vert}{r_{\sA}+r_{\sB}-\vert \bx_{\sB} -\bx_{\sA} \vert}\right). 
\ee
Substituting for ${\cal T}^{(1)}$ from (\ref{T1}) into (\ref{qn}) written for $n=1$, and then using (\ref{derR}), it is easily seen that
\be \label{q1f}
q_1=\frac{(1+\gamma) r_c}{1+\bn_{\sA}.\bn_{\sB}}\left( \frac{1}{r_{\sA}}+\frac{1}{r_{\sB}}\right).
\ee
Taking into account this determination of $q_1$, it would be possible to carry out the calculation of ${\cal T}^{(2)}$ via (\ref{U2}). Then, $q_2$ could be derived from (\ref{qn}) taken for $n=2$. Consequently, ${\cal T}^{(3)}$ could be deduced from (\ref{U3}). However, we shall see in the next section that the method can be simplified by making use of the differential equation governing the variation of the angular coordinate along the light ray. In particular, it turns out that determining $q_2$ is not indispensable for calculating ${\cal T}^{(3)}$.
 
\section{Simplification of the procedure by using a constraint equation} \label{pract_impl}

Equations (\ref{ip2}) and (\ref{eqdf2}) are equivalent to the geodesic equations
\begin{eqnarray} 
\frac{d\varphi}{d\zeta}=\frac{J}{r^2 {\cal U}(r)}, \label{df} \\
\frac{dr}{d\zeta}=\varepsilon \frac{E}{r {\cal U}(r)}\sqrt{r^2{\cal U}(r)-b^2} .\label{dr}
\end{eqnarray}
Eliminating the affine parameter $\zeta$ between (\ref{df}) and (\ref{dr}) leads to
\be \label{dfdr}
\frac{d\varphi}{dr}=\varepsilon\frac{b}{r}\frac{1}{\sqrt{r^2{\cal U}(r)-b^2}}.
\ee

Since $\varepsilon=1$ when conditions (\ref{rinc}) and (\ref{cdc}) are met, integrating (\ref{dfdr}) and taking into account (\ref{phiAB}) yield thereby
\be \label{fab}
\arccos\mu=\int_{r_{\sA}}^{r_{\sB}}\frac{b}{r\sqrt{r^2{\cal U}(r)-b^2}}dr.
\ee
Equation (\ref{fab}) implicitly determines $b$ as a function of $\bx_{\sA}$ and $\bx_{\sB}$. So it may be expected that this equation implies some conditions on the coefficients $q_n$ which may be used to simplify the calculations. 

Replacing ${\cal U}$ by (\ref{invAB}) and $b$ by (\ref{expb}) into (\ref{fab}), it may be seen that  
\be \label{expfab}
\arccos\mu=r_c\int_{r_{\sA}}^{r_{\sB}}\frac{dr}{rs}+\frac{1}{r_c}\sum_{n=1}^{\infty}\left( \frac{m}{r_c}\right)^n \int_{r_A}^{r_B}W_ndr,
\ee
where the $W_n$'s are functions of $r$ which may be written in the form 
\begin{eqnarray}  
W_1= -(1+\gamma)\frac{r_c^3 }{s^3}+q_1\frac{r_c^2 r}{s^3}, \label{Y1} \\
W_n=\sum_{k=0}^{3(n-1)} W_{kn}(q_1,\dots ,q_{n-1})r_c^{3n-k} 
\frac{r^{k-n+1}}{s^{2n+1}} +q_n\frac{r_c^2 r}{s^3} \label{Yn}
\end{eqnarray}
for $n\geq 2$, with the terms $W_{kn}(q_1,\dots,q_{n-1})$ being polynomials in $q_1,\dots,q_{n-1}$. Taking into account (\ref{fab0}), it is immediately seen that (\ref{expfab}) reduces to
\be \label{expf0}
\sum_{n=1}^{\infty}\left( \frac{m}{r_c}\right)^n \int_{r_A}^{r_B}W_n dr=0.
\ee
Since (\ref{expf0}) holds whatever $m$, it is clear that (\ref{expfab}) is equivalent to the infinite set of equations
\be \label{setY}
\int_{r_{\sA}}^{r_{\sB}}W_n dr=0, \qquad n=1,2,\dots 
\ee

The set of constraint equations (\ref{setY}) may be systematically used for simplifying our problem. Let us consider the functions $U_n^{\ast}$ defined as 
\begin{eqnarray} 
U_1^{\ast}=U_1, \label{R1} \\
U_n^{\ast}=U_n+\sum_{p=1}^{n-1}k_{pn}W_p \label{Rn}
\end{eqnarray}
for $n\geq 2$, where the $k_{pn}$'s are arbitrary quantities which do not depend on $r$. Taking into account (\ref{setY}), it is immediately seen that
\be \label{XRn}
\int_{r_{\sA}}^{r_{\sB}}U_n dr=\int_{r_{\sA}}^{r_{\sB}}U_n^{\ast}dr.
\ee
Hence ${\cal T}^{(n)}$ may be rewritten in the form
\be \label{tRn}
{\cal T}^{(n)}(\bx_{\sA},\bx_{\sB}) = \frac{1}{c}\left( \frac{m}{r_c}\right)^n \int_{r_{\sA}}^{r_{\sB}}U_n^{\ast} dr .
\ee
Of course, the remark formulated just after (\ref{TnAB}) might be reproduced here. 

It is easily seen that a judicious choice of the quantities $k_{pn}$ enables to shorten the expressions involved in (\ref{tRn}) when $n\geq2$. Until $n=3$, only the expression of $q_1$ is needed. Indeed, for $n=2$, setting $k_{12}=\frac{1}{2} q_1$ removes the term in $q_1^2$ and leads to
\be \label{R2}
\fl U_2^{\ast}=-\frac{\kappa r_c^4}{rs^3}+\frac{(1+\gamma)q_1 r_c^3}{2s^3}+\frac{[2\kappa-(1+\gamma)^2]r_c^2r}{2s^3}.
\ee 
For $n=3$, choosing $k_{13}=q_2$ and $k_{23}=0$ remove the terms involving $q_2$. Then $U_3^{\ast}$ reduces to
\begin{eqnarray} 
\fl U_3^{\ast}=\frac{\kappa_3 r_c^7}{r^2s^5} -\frac{\kappa q_1 r_c^6}{rs^5}- \frac{[2\kappa_3-(1+\gamma)(\kappa +q_1^2)]r_c^5}{s^5} \nonumber \\
+\frac{[2\kappa - 3(1+\gamma)^2-q_1^2]q_1r_c^4r}{2s^5} \nonumber \\
+\frac{[2\kappa_3-(1+\gamma)(2\kappa - q_1^2)+(1+\gamma)^3]r_c^3r^2}{2s^5}. \label{R3}
\end{eqnarray}

It is thus proved that owing to the constraint equation (\ref{fab}), only the determination of $q_1$ is required for calculating the functions ${\cal T}^{(2)}$ and ${\cal T}^{(3)}$.

{\it Remark}. It may be pointed out that the coefficients $q_n$ could be directly inferred from the constraint equation without differentiating the functions ${\cal T}^{(n)}$ with respect to $\mu$. Indeed, it follows from (\ref{Y1}), (\ref{Yn}) and (\ref{setY}) that
\begin{eqnarray}
\fl q_1=\frac{1+\gamma}{r_c}\frac{r_{\sA}\sqrt{r_{\sB}^{2}-r_{c}^{2}}-r_{\sB}\sqrt{r_{\sA}^{2}-r_{c}^{2}}}
{\sqrt{r_{\sB}^{2}-r_{c}^{2}}-\sqrt{r_{\sA}^{2}-r_{c}^{2}}}, \label{q1} \\
\fl q_n=-\frac{1}{r_c}\frac{\sqrt{r_{\sA}^{2}-r_{c}^{2}}\,\sqrt{r_{\sB}^{2}-r_{c}^{2}}}{\sqrt{r_{\sB}^{2}-r_{c}^{2}}-\sqrt{r_{\sA}^{2}-r_{c}^{2}}}\nonumber \\
\times\sum_{k=0}^{3(n-1)} W_{kn}(q_1,\dots ,q_{n-1})r_{c}^{3n-k-1}\int_{r_{\sA}}^{r_{\sB}} 
\frac{r^{k-n+1}}{s^{2n+1}}dr \label{qnc}
\end{eqnarray}
for $n\geq2$. Equation (\ref{qnc}) shows that $q_n$ can be determined once $q_1, \dots,q_{n-1}$ are known. 

It is easily checked that (\ref{q1}) is equivalent to (\ref{q1f}). Indeed, noting that
\be \label{RAB}
\sqrt{r_{\sB}^{2}-r_{c}^{2}}-\sqrt{r_{\sA}^{2}-r_{c}^{2}}=\vert \bx_{\sB}-\bx_{\sA} \vert,
\ee
and then taking into account (\ref{rc}), (\ref{NnA}) and (\ref{NnB}), it may be seen that (\ref{q1}) transforms into
\be \label{q1a}
q_1=(1+\gamma)\frac{\bN_{\sAB}.\bn_{\sB}-\bN_{\sAB}.\bn_{\sA}}{\vert \bn_{\sA}\times\bn_{\sB}\vert}.
\ee
Substituting $r_{\sA}\bn_{\sA}$ for $\bx_{\sA}$ and $r_{\sB}\bn_{\sB}$ for $\bx_{\sB}$ into the numerator of  the right-handside of (\ref{NAB}) yields
\be \label{Nnn}
\bN_{\sAB}.\bn_{\sB}-\bN_{\sAB}.\bn_{\sA}=\frac{(r_{\sA}+r_{\sB})(1-\bn_{\sA}.\bn_{\sB})}{\vert\bx_{\sB}-\bx_{\sA}\vert}.
\ee
Finally, substituting for $\bN_{\sAB}.\bn_{\sB}-\bN_{\sAB}.\bn_{\sA}$ from (\ref{Nnn}) into (\ref{q1a}), and then noting that (\ref{rc}) is equivalent to
\[
\frac{1}{\vert\bx_{\sB}-\bx_{\sA}\vert}=\frac{r_c}{r_{\sA} r_{\sB}} \frac{1}{\vert \bn_{\sA}\times\bn_{\sB}\vert},
\] 
it is immediately seen that (\ref{q1f}) is recovered.

\section{Time transfer function up to the third order} \label{TTF3rd}

We are now in a position to determine the perturbation terms involved in the expansion of the time transfer function up to the order $G^3$. The term ${\cal T}^{(1)}$ has been already treated in section \ref{princmeth}. For $n= 2$ and $n= 3$, it follows from (\ref{R2}) and (\ref{R3}) that ${\cal T}^{(n)}$ may be written in the form
\be \label{intp}
\fl{\cal T}^{(n)}(\bx_{\sA}, \bx_{\sB})=\frac{1}{c}\left(\frac{m}{r_c}\right)^n\sum_{k=0}^{\sigma(n)}U^{\ast}_{kn}(q_1)r_c^{3n-k-2}\int_{r_{\sA}}^{r_{\sB}}\frac{r^{k-n+1}}{s^{2n-1}}\,dr
\ee
where $\sigma(2)=2$ and $\sigma(3)=4$, with the coefficients $U^{\ast}_{kn}$ being polynomials in $q_1$. The integrals occurring into the right-hand side of (\ref{intp}) are elementary and can be expressed in terms of $r_{\sA}$, $r_{\sB}$, $r_c$, $\sqrt{r_{\sA}^2-r_c^2}$ and $\sqrt{r_{\sB}^2-r_c^2}$. For the explicit calculations, it is convenient to write (\ref{NnA}) and (\ref{NnB}) in the form
\numparts
\begin{eqnarray} 
\sqrt{r_{\sA}^{2}-r_{c}^{2}}=\frac{r_{\sA}(r_{\sB}\mu -r_{\sA})}{\vert \bx_{\sB} -\bx_{\sA} \vert}, \label{Nn1A} \\ \sqrt{r_{\sB}^{2}-r_{c}^{2}}=\frac{r_{\sB}(r_{\sB} -r_{\sA}\mu)}{\vert \bx_{\sB} -\bx_{\sA} \vert}. \label{Nn1B}
\end{eqnarray}
\endnumparts
Using (\ref{rc}), (\ref{q1f}), (\ref{Nn1A}) and (\ref{Nn1B}), it may be seen that ${\cal T}^{(2)}$ and ${\cal T}^{(3)}$ can be expressed in terms of $r_{\sA}r_{\sB}$, $1/r_{\sA}+1/r_{\sB}$, $ \vert \bx_{\sB} -\bx_{\sA} \vert$ and $\mu$. It has been already emphasized that the explicit calculations can be performed with any symbolic computer program. Of course, a simple hand calculation is also possible. For $n=2$, the result is straightforwardly obtained. For $n=3$, however, the calculations are somewhat lengthy and tedious. For this reason, some hints concerning this case are delivered in the appendix. 

We have seen in section \ref{princmeth} that the expressions thus obtained can be considered as valid even when conditions (\ref{rinc}) and (\ref{cdc}) are not fulfilled. So we can formulate the following proposition.

\begin{proposition} \label{propT3}
Let $\bx_{\sA}$ and $\bx_{\sB}$ be two points in ${\cal D}_h$ such that both the conditions $\bn_{\sA}\neq\bn_{\sB}$ and (\ref{strl}) are met. Then ${\cal T}^{(2)}$ and ${\cal T}^{3)}$ are given by
\begin{eqnarray} 
\fl{\cal T}^{(2)}( \bx_{\sA}, \bx_{\sB}) = \frac{m^2}{r_{\sA}r_{\sB}} \frac{\vert \bx_{\sB} -\bx_{\sA} \vert}{c}\bigg[ \kappa\frac{\arccos \bn_{\sA}.\bn_{\sB}}{\vert\bn_{\sA}\times\bn_{\sB}\vert}-
\frac{(1+\gamma)^2}{1+\bn_{\sA}.\bn_{\sB}}\bigg], \label{T2} \\
\fl{\cal T}^{(3)}(\bx_{\sA},\bx_{\sB})=\frac{m^3}{r_{\sA}r_{\sB}}\left( \frac{1}{r_{\sA}}+\frac{1}{r_{\sB}}\right)\frac{\vert\bx_{\sB}-\bx_{\sA}\vert}{c(1+\bn_{\sA}.\bn_{\sB})} \bigg\lbrack \kappa_3 -(1+\gamma)\kappa\frac{\arccos \bn_{\sA} .\bn_{\sB}}{\vert\bn_{\sA}\times\bn_{\sB}\vert}  \nonumber \\
+\frac{(1+\gamma )^3}{1+\bn_{\sA}.\bn_{\sB}}\bigg\rbrack, \label{T3}
\end{eqnarray}
where the coefficients $\kappa$ and $\kappa_3$ are determined by (\ref{kappa}) and (\ref{kappa3}), respectively.
\end{proposition}

The formula obtained for ${\cal T}^{(3)}$ is new\footnote{We have given the expression of ${\cal T}^{(3)}$ without demonstration in \cite{teyssandier3}.}. Concerning ${\cal T}^{(2)}$, it appears that (\ref{T2}) coincides with the expression previously obtained by completely different methods in \cite{leponcin1}, \cite{teyssandier1} and \cite{ashby}\footnote{See also \cite{klioner1}, where the expression of ${\cal T}^{(2)}$ is obtained in harmonic coordinates by an integration of the null geodesic equations.}. This concordance confirms the reliability of the procedure presented here.

In the case of a radial null geodesic, it is immediately inferred from 
\be \label{lim2}
\lim_{\bn_{\sB}\rightarrow \bn_{\sA}}\frac{\arccos \bn_{\sA}.\bn_{\sB}}{\vert\bn_{\sA}\times\bn_{\sB}\vert}=1
\ee 
that the system (\ref{Tr1})-(\ref{Tr3}) is recovered from (\ref{T1}), (\ref{T2}) and (\ref{T3}). This agreement constitutes another confirmation of the validity of the new procedure.

In general relativity, the expressions of ${\cal T}^{(1)}$, ${\cal T}^{(2)}$ and ${\cal T}^{(3)}$ are obtained by setting $\gamma=1$, $\kappa=\frac{15}{4}$ and $\kappa_3=\frac{9}{2}$.

\section{Enhanced terms in ${\cal T}^{(1)}$, ${\cal T}^{(2)}$ and ${\cal T}^{(3)}$} \label{Suit3rd}

In the present work, the time transfer function ${\cal T}$ is obtained in the form of an asymptotic expansion in power series in $G$ (or $m$) provided that condition (\ref{strl}) is met. However, it is clear that the physical reliability of this expansion requires that inequalities as follow 
\be \label{CTn}
\left\vert{\cal T}^{(n)}( \bx_{\sA}, \bx_{\sB})\right\vert \ll \left\vert{\cal T}^{(n-1)}( \bx_{\sA}, \bx_{\sB})\right\vert 
\ee
are satisfied for any $n\geq 1$, with ${\cal T}^{(0)}( \bx_{\sA}, \bx_{\sB})$ being conventionally defined as  
\[
{\cal T}^{(0)}( \bx_{\sA}, \bx_{\sB})=\frac{1}{c}\vert \bx_{\sB}-\bx_{\sA}\vert.
\]

The results obtained in the previous section enable us to find the conditions ensuring inequalities (\ref{CTn}) for $n=1, 2, 3$. It is clear that the magnitude of the functions given by (\ref{T1}), (\ref{T2}) and (\ref{T3}) may be extremely large when points $\bx_{\sA}$ and $\bx_{\sB}$ are located in almost opposite directions. This behaviour corresponds to the `enhanced terms' determined up to $G^2$ for the light deflection in \cite{klioner1} and up to $G^3$ for the time transfer function in \cite{ashby}. Indeed, it is straightforwardly derived from (\ref{rc}) that
\be \label{asyn}
\frac{1}{1+\bn_{\sA}.\bn_{\sB}}\sim \frac{2 r_{\sA}^2 r_{\sB}^2}{(r_{\sA} +r_{\sB})^2}\frac{1}{r_c^2}
\ee
when $1+\bn_{\sA}.\bn_{\sB}\rightarrow 0$. Using this relation to eliminate $1+\bn_{\sA}.\bn_{\sB}$, the following proposition is easily deduced from (\ref{T1}), (\ref{T2}) and (\ref{T3}). 

\begin{proposition} \label{propAsymp}
When $\bx_{\sA}$ and $\bx_{\sB}$ tend to be located in opposite directions ({\it i.e.} $1+\bn_{\sA}.\bn_{\sB}\rightarrow 0$), the first three perturbation terms in the time transfer function are enhanced according to the asymptotic expressions
\begin{eqnarray}
{\cal T}^{(1)}_{enh}(\bx_{\sA}, \bx_{\sB})\sim \frac{(1+\gamma)m}{c}\ln\left(\frac{4 r_{\sA} r_{\sB}}{r^2_c}\right), \label{asT1}\\
{\cal T}^{(2)}_{enh}(\bx_{\sA}, \bx_{\sB})\sim -2 \frac{(1+\gamma)^2 m^2}{c(r_{\sA}+r_{\sB})} \frac{r_{\sA} r_{\sB}}{r^2_c}, \label{asT2}\\
{\cal T}^{(3)}_{enh}(\bx_{\sA}, \bx_{\sB})\sim 4\frac{(1+\gamma)^3 m^3}{c(r_{\sA}+r_{\sB})^2} \left(\frac{r_{\sA} r_{\sB}}{r^2_c}\right)^2. \label{asT3}
\end{eqnarray}
\end{proposition}

These expressions confirm the formulae obtained in \cite{ashby} by a different method. It is worthy noticing that, at least up to $G^3$, $\gamma$ is the only post-Newtonian parameter involved in the enhanced terms. When $\bx_{\sA}$ and $\bx_{\sB}$ tend to be located in opposite directions, the asymptotic behavior of each function ${\cal T}^{(n)}_{enh}$ is such that 
\be \label{CasTn}
\left\vert{\cal T}^{(n)}_{enh}(\bx_{\sA}, \bx_{\sB})\right\vert \lesssim k_n \frac{2(1+\gamma) m}{r_{\sA}+r_{\sB}} \frac{r_{\sA} r_{\sB}}{r^2_c} \left\vert{\cal T}^{(n-1)}_{enh}(\bx_{\sA}, \bx_{\sB})\right\vert
\ee
for $n=1,2,3$, with $k_1=2$, $k_2=k_3=1$ and ${\cal T}^{(0)}_{enh}(\bx_{\sA}, \bx_{\sB})\sim r_{\sA}+r_{\sB}$. For $n=3$, the formula (\ref{CasTn}) is straightforwardly derived from (\ref{asT2}) and (\ref{asT3}) (the symbol $\lesssim$ could be replaced by $\sim$). For $n=1$, the formula results from the fact that $\ln x<x$  for any $x>0$. Lastly, for $n=2$, (\ref{CasTn}) obviously follows from the fact that $\ln (4 r_{\sA} r_{\sB}/r^2_c) \rightarrow \infty$ when $1+\bn_{\sA}.\bn_{\sB}\rightarrow 0$.

It results from (\ref{CasTn}) that inequalities (\ref{CTn}) are satisfied for $n=1,2,3$ as long as the zeroth-order distance of closest approach $r_c$ is such that a condition as follows
\be \label{asyp}
\frac{2 m}{r_{\sA}+r_{\sB}} \frac{r_{\sA} r_{\sB}}{r^2_c} \ll 1
\ee
is fulfilled. This inequality coincides with the condition ensuring the validity of the asymptotic expansions obtained in \cite{ashby}. It may be expected that (\ref{asyp}) is sufficient to ensure inequality (\ref{CTn}) at any order. 

Combined with (\ref{asyn}), (\ref{asyp}) means that our results are reliable when $\bx_{\sA}$ and $\bx_{\sB}$ tend to be located in opposite directions as long as an inequality as follows 
\be \label{vasy}
\pi - \arccos \bn_{\sA}.\bn_{\sB} \gg \sqrt{\frac{2m(r_{\sA}+r_{\sB})}{r_{\sA} r_{\sB}}}
\ee
is satisfied. Such a condition clearly indicates that our procedure is not appropriate for the case of a gravitational lensing configuration.

\section{Application to some solar system experiments} \label{Apexp}

Condition (\ref{asyp}) is fulfilled in experiments performed with photons exchanged between a spacecraft in the outer solar system and a ground station. Indeed, noting that 
\[
\frac{m}{r_c} \frac{r_{\sB}}{r_c}<\frac{2 m}{r_{\sA}+r_{\sB}} \frac{r_{\sA} r_{\sB}}{r^2_c}<2\frac{m}{r_c} \frac{r_{\sB}}{r_c}
\]
holds if $r_{\sA}>r_{\sB}$, replacing $m$ by half the Schwarzschild radius of the Sun, $m_{\odot}$, and then putting $r_{\sB}=$ 1 au, we find that inequalities
\be \label{exasyp}
4.56 \times 10^{-4} \times \frac{R_{\odot}^2}{r_c^2}<\frac{2 m_{\odot}}{r_{\sA}+r_{\sB}} \frac{r_{\sA} r_{\sB}}{r^2_c}<
9.12\times 10^{-4} \times \frac{R_{\odot}^2}{r_c^2}
\ee
hold if $r_{\sA}>r_{\sB}$, with $R_{\odot}$ denoting the radius of the Sun. We put $R_{\odot}=6.96\times 10^8$ m. The other numerical parameters of the Sun used throughout this section are taken from \cite{iers}.  

The formulae (\ref{asT1})-(\ref{asT3}) enable us to discuss the relevance of the terms ${\cal T}^{(2)}_{enh}$ and ${\cal T}^{(3)}_{enh}$ in a proposed mission like SAGAS, for instance. Indeed, this project plans to measure the parameter $\gamma$ up to an accuracy reaching $10^{-8}$ with light rays travelling between a spacecraft moving in the outer solar system and the Earth. For $r_{\sA}=50$ au and $r_{\sB}=1$ au, the travel time of a ray passing in close proximity to the Sun (conjunction) is about $2.54\times 10^4$ s. It follows from (\ref{asT1}) that ${\cal T}^{(1)}$ is decreasing from 158 $\mu$s to 126 $\mu$s when $r_c$ varies from $R_{\odot}$ to $5 R_{\odot}$. As a consequence, reaching an accuracy of $10^{-8}$ on the measurement of $\gamma$ requires to determine the light travel time with an accuracy of 0.7 ps. The numerical values of the respective contributions of ${\cal T}^{(2)}_{enh}$ and ${\cal T}^{(3)}_{enh}$ are indicated in table \ref{table1}. It is clear that the contribution of the enhanced term of order $G^3$ is larger than 2 ps when $r_c < 2R_{\odot}$. The same order of magnitude for ${\cal T}^{(3)}_{enh}$ may be expected in other proposed missions like ODYSSEY, LATOR or ASTROD. 

The above discussion also reveals that an experiment like SAGAS would enable to determine the post-post-Newtonian parameter $\kappa$ with a relative precision amounting to $7 \times 10^{-3}$. In the solar system, indeed, the term proportional to $\kappa$ in (\ref{T2}) yields the asymptotic contribution
\be \label{asT2k}
{\cal T}^{(2)}_{\kappa}(\bx_{\sA}, \bx_{\sB}) \sim \frac{\kappa \pi m_{\odot}^2}{c r_c}
\ee
when (\ref{asyn}) holds. For a ray grazing the Sun ($r_c=R_{\odot}$), one has ${\cal T}^{(2)}_{\kappa}\approx 123$ ps if $\kappa =15/4$. Hence the conclusion.

Before closing this study, it is worthy of note that the first-order contribution ${\cal T}^{(1)}_{S}$ to the time transfer function due to the gravitomagnetic effect of the solar rotation may be compared with the third-order enhanced term. Indeed, it is easily inferred from equation (62) in \cite{linet} that for a ray travelling in the equatorial plane of the Sun
\be \label{asT1S}
\left\vert{\cal T}^{(1)}_{S}(\bx_{\sA}, \bx_{\sB})\right\vert \sim \frac{2(1+\gamma)G S_{\odot}}{c^4 r_c}
\ee
when (\ref{asyn}) is checked, with $S_{\odot}$ being the angular momentum of the Sun. According to helioseismology, we can take $S_{\odot} \approx 2\times 10^{41}$ kg m$^2$ s$^{-1}$ (see, e.g., \cite{komm}). So, in the case where $r_c=R_{\odot}$, we have $\vert{\cal T}^{(1)}_{S}(\bx_{\sA}, \bx_{\sB})\vert\approx 10$ ps. Furthermore, the contribution ${\cal T}^{(1)}_{J_2}$ due to the solar quadrupole moment $J_{2\odot}$ must also be considered for rays grazing the Sun. Using equation (24) in \cite{leponcin2} for a ray travelling in the equatorial plane gives
\be \label{asJ2}
{\cal T}^{(1)}_{J_2}(\bx_{\sA}, \bx_{\sB}) \sim \frac{(1+\gamma)m_{\odot}}{c} J_{2\odot} \frac{R_{\odot}^2}{r_c^2}.
\ee
Taking $J_{2\odot} \approx2\times 10^{-7}$ and putting $r_c=R_{\odot}$, (\ref{asJ2}) leads to ${\cal T}^{(1)}_{J_2}\approx2$ ps\footnote{It may be pointed out that the time delay due to the cosmological constant is much smaller than the contribution of ${\cal T}^{(3)}$ in any solar system experiment since $\Lambda r_{\sA}^3/9 < 10^{-22}$ s for $r_{\sA}<100$ au with $\Lambda \approx 10^{-52}$ m$^{-2}$.}. 

\begin{table}
\caption{\label{table1}Numerical values in ps of the main stationary contributions to the light travel time in the solar system for various values of $r_c/R_{\odot}$. In each case, $r_{\sA}=50$ au and $r_{\sB}=1$ au. The parameters $\gamma$ and $\kappa$ are taken as $\gamma=1$ and $\kappa=15/4$, respectively. For the numerical estimates of $\vert{\cal T}^{(1)}_{S}\vert$ and ${\cal T}^{(1)}_{J_2}$, the light ray is assumed to propagate in the equatorial plane of the Sun. The dynamical effects due to the planetary perturbations are not taken into account.} 

\begin{indented}
\lineup
\item[]\begin{tabular}{@{}*{6}{l}}
\br                              
\0\0$r_c/R_{\odot}$&\m$\left\vert{\cal T}^{(1)}_{S}\right\vert$&\m${\cal T}^{(1)}_{J_2}$&\m\0\0${\cal T}^{(2)}_{enh}$&\m\0${\cal T}^{(2)}_{\kappa}$&\m${\cal T}^{(3)}_{enh}$\cr 
\mr
\0\0\0\0$1$&\m\m 10&\m \02&\m$-17616$&\m 123 &\m 31.5\cr
\0\0\0\0$2$&\m\m \05&\m \00.5&\m\0$-4404$&\m \061.5 &\m \02 \cr 
\0\0\0\0$5$ &\m\m \02 &\m \00.08&\m\0\0$-704.6$&\m \024.6   &\m \0$0.05$  \cr 
\br
\end{tabular}
\end{indented}
\end{table}

\section{Impact parameter up to the third order} \label{imp_par3}
  
The coefficient $q_1$ has been previously inferred from (\ref{T1}) in section \ref{princmeth}. Substituting for  ${\cal T}^{(2)}$ from (\ref{T2}) into (\ref{qn}), and then taking into account that $r_c \vert \bx_{\sB}-\bx_{\sA}\vert/r_{\sA} r_{\sB} = \sqrt{1-\mu^2}$, we get
\be \label{q_2}
\fl q_2=\kappa -\frac{1}{\vert \bx_{\sB}-\bx_{\sA}\vert^2}\bigg\lbrace \kappa\,\frac{\arccos \mu}{\sqrt{1-\mu^2}} Q_{\sAB}
+\frac{(1+\gamma)^2}{1+\mu}\left[\vert \bx_{\sB}-\bx_{\sA}\vert^2-Q_{\sAB}\right]  \bigg\rbrace,
\ee
where $Q_{\sAB}$ is defined as
\be \label{QAB}
Q_{\sAB}=\vert \bx_{\sB}-\bx_{\sA}\vert^2\mu - r_{\sA}r_{\sB}(1-\mu^2).
\ee
Then, substituting  for ${\cal T}^{(3)}$ from (\ref{T3}) into (\ref{qn}) and using (\ref{QAB}), we obtain
\begin{eqnarray}  
\fl q_3=\frac{r_{\sA}+r_{\sB}}{\vert \bx_{\sB}-\bx_{\sA}\vert^{3}}\frac{\sqrt{1-\mu}}{\sqrt{1+\mu}}\, \bigg\lbrace \kappa_3 \left[\vert \bx_{\sB}-\bx_{\sA}\vert^2-Q_{\sAB}\right] \nonumber \\
-(1+\gamma )\kappa \bigg\lbrack \vert \bx_{\sB}-\bx_{\sA}\vert^2+
\frac{\vert \bx_{\sB}-\bx_{\sA}\vert^2(1-\mu)-Q_{\sAB}}{\sqrt{1-\mu^2}} 
\arccos \mu\bigg\rbrack \nonumber \\
+(1+\gamma )^3\frac{\vert \bx_{\sB}-\bx_{\sA}\vert^2(2-\mu)-Q_{\sAB}}{1+\mu} \bigg\rbrace . \label{q_3}
\end{eqnarray}

The dimensionless coefficients $q_n$ can be expressed in terms of the sine (or cosine) of the angles formed by $\bn_{\sA}$, $\bn_{\sB}$ and $\bN_{\!\sA\sB}$. Noting that 
\be \label{iden1}
r_c\left(\frac{1}{r_{\sA}}+\frac{1}{r_{\sB}}\right)=\vert\bN_{\!\sA\sB} \times \bn_{\sA}\vert +\vert\bN_{\!\sA\sB} \times \bn_{\sB}\vert 
\ee
and that (\ref{QAB}) may be written in the form
\be \label{iden2}
Q_{\sAB} = \vert \bx_{\sB}-\bx_{\sA}\vert^2 (\bN_{\! \sA\sB}.\bn_{\sA})(\bN_{\! \sA\sB}.\bn_{\sB}),
\ee
a proposition as follows is straightforwardly inferred from (\ref{q1f}), (\ref{q_2}) and (\ref{q_3}).

\begin{proposition} \label{propq123}
Under the assumption of proposition \ref{propT3}, the coefficients $q_1$, $q_2$ and $q_3$ involved in the expansion of the impact parameter $b$ of a quasi-Minkowskian light ray joining $\bx_{\sA}$ and $\bx_{\sB}$ are given by 
\begin{eqnarray} 
\fl q_1 (\bx_{\sA} ,\bx_{\sB} )= ( 1+\gamma ) \frac{\vert\bN_{\!\sA\sB} \times \bn_{\sA}\vert + \vert\bN_{\!\sA\sB} \times \bn_{\sB}\vert}{1+\bn_{\sA}.\bn_{\sB}},  \label{q_1N} \\
\fl q_2(\bx_{\sA},\bx_{\sB})=\kappa \left[ 1-\frac{(\bN_{\! \sA\sB}.\bn_{\sA})(\bN_{\! \sA\sB}.\bn_{\sB})}{\vert\bn_{\sA}\times\bn_{\sB}\vert}\arccos \bn_{\sA}.\bn_{\sB}\right] \nonumber \\
-(1+\gamma )^2\frac{1-(\bN_{\! \sA\sB}.\bn_{\sA})(\bN_{\! \sA\sB}.\bn_{\sB})}{1+\bn_{\sA}.\bn_{\sB}}, \label{q_2N}\\
\fl q_3 (\bx_{\sA} ,\bx_{\sB})=\frac{\vert\bN_{\!\sA\sB} \times \bn_{\sA}\vert + \vert\bN_{\!\sA\sB} \times \bn_{\sB}\vert}{1+\bn_{\sA}.\bn_{\sB}} \bigg\lbrace\kappa_3\left[ 1-(\bN_{\! \sA\sB}.\bn_{\sA})(\bN_{\! \sA\sB}.\bn_{\sB}) \right]\nonumber \\
-(1+\gamma )\kappa \left[ 1 +\frac{1-\bn_{\sA}.\bn_{\sB} -(\bN_{\! \sA\sB}.\bn_A)(\bN_{\! \sA\sB}.\bn_{\sB})}{\vert\bn_{\sA}\times\bn_{\sB}\vert} 
\arccos \bn_{\sA}.\bn_{\sB}\right] \nonumber \\
+(1+\gamma )^3\frac{2-\bn_{\sA}.\bn_{\sB} -(\bN_{\! \sA\sB}.\bn_{\sA})(\bN_{\! \sA\sB}.\bn_{\sB})}{1+\bn_{\sA}.\bn_{\sB}} \bigg\rbrace . \label{q_3N}
\end{eqnarray}
\end{proposition}

Equations (\ref{q_1N}) and (\ref{q_2N}) are identical to the expressions obtained in \cite{teyssandier2}. We remark that the first-order expression of $b$ yielded by (\ref{expb}) and (\ref{q_1N}) coincides with the Euclidean norm of the vector $\bm d$' given by equation (62) in \cite{klioner1}. On the other hand, the formula (\ref{q_3N}) is new and completes the implementation of our method up to the third order.

To finish, it may be easily seen that condition (\ref{asyp}) ensures that
\be \label{asqn}
\vert q_n\vert \ll \vert q_{n-1}\vert
\ee  
holds for $n=1, 2, 3$, with $q_0$ being conventionally defined by $q_0=1$.

\section{Case of a light ray emitted at infinity} \label{rinfin}

A quasi-Minkowskian light ray coming from infinity in a initial direction defined by a given unit vector $\bN_{e}$ and observed at a given point $\bx_{\sB}$ is a relevant limiting case for modelling a lot of astrometric measurements. According to a notation introduced in \cite{teyssandier2}, such a ray is denoted by $\Gamma_s(\bN_{e},\bx_{\sB})$. The corresponding null geodesic is assumed to be a perturbation in powers of $G$ of the straight segment defined by the parametric equations
\be \label{param}
\fl x^{0}_{(0)}(\lambda)=ct_{\sB}+\lambda r_c , \quad \bx_{(0)}(\lambda)=\lambda r_c\bN_e+\bx_{\sB}, \quad -\infty <\lambda\leq 0, 
\ee
where
\be \label{rcinf}
r_c=r_{\sB}\vert\bN_{e}\times\bn_{\sB}\vert .
\ee
 
In order to ensure that condition (\ref{strl}) is satisfied for any point $\bx_{\sA}$ of $\Gamma_s(\bN_{e},\bx_{\sB})$, $\bN_{e}$ and $\bx_{\sB}$ must be supposed to satisfy the condition
\be \label{stli}
\vert \lambda r_c\bN_e+\bx_{\sB} \vert>r_h \quad \mbox{when} -\infty <\lambda\leq 0.
\ee
This condition means that the straight segment coming from infinity in the direction $\bN_{e}$ and ending at $\bx_{\sB}$ is entirely lying in ${\cal D}_h$.  

Then the following proposition can be stated.

\begin{proposition} \label{propqinf}
Let $\bN_{e}$ be a unit vector and $\bx_{\sB}$ a point in ${\cal D}_h$ fulfilling condition (\ref{stli}). The impact parameter of a quasi-Minkowskian light ray emitted at infinity in the direction $\bN_{e}$ and arriving at $\bx_{\sB}$ is given by expansion (\ref{expb}), where $r_c$ is expressed by (\ref{rcinf}) and the coefficients $q_1$, $q_2$ and $q_3$ are yielded by
\begin{eqnarray} 
\fl q_1(\bN_{e},\bx_{\sB})= ( 1+\gamma ) \frac{\vert\bN_{e}\times\bn_{\sB}\vert}{1-\bN_{e}.\bn_{\sB}},  \label{q1inf} \\
\fl q_2(\bN_{e},\bx_{\sB})=\kappa \left[ 1+(\bN_{e}.\bn_{\sB})\frac{\pi-\arccos \bN_{e}.\bn_{\sB}}{\vert\bN_{e}\times\bn_{\sB}\vert}\right]-(1+\gamma )^2\frac{1+\bN_{e}.\bn_{\sB}}{1-\bN_{e}.\bn_{\sB}}, \label{q2inf}\\
\fl q_3 (\bN_{e},\bx_{\sB})=\frac{\vert\bN_{e}\times\bn_{\sB}\vert}{1-\bN_{e}.\bn_{\sB}} \bigg\lbrace \kappa_3\left( 1+\bN_{e}.\bn_{\sB}\right)+2(1+\gamma )^3\frac{1+\bN_{e}.\bn_{\sB}}{1-\bN_{e}.\bn_{\sB}} \nonumber \\
-(1+\gamma)\kappa\left[1+\left(1+2\bN_{e}.\bn_{\sB} \right)  
\frac{\pi-\arccos \bN_{e}.\bn_{\sB}}{\vert\bN_{e}\times\bn_{\sB}\vert} \right] \bigg\rbrace . \label{q3inf}
\end{eqnarray} 
\end{proposition}

{\bf Proof of proposition \ref{propqinf}}  Let $\bx_{\sA}$ be a point lying on $\Gamma_s(\bN_{e},\bx_{\sB})$. It is 
clear that the part of $\Gamma_s(\bN_{e},\bx_{\sB})$ joining $\bx_{\sA}$ and $\bx_{\sB}$ coincides with a quasi-Minkowskian null geodesic path $\Gamma_s(\bx_{\sA},\bx_{\sB})$. So, the impact parameters of $\Gamma_s(\bN_{e},\bx_{\sB})$ and $\Gamma_s(\bx_{\sA},\bx_{\sB})$ are equal. As a consequence, the coefficients $q_1, q_2$ and $ q_3$ can be obtained as functions of $\bN_{e}$ and $\bx_{\sB}$ by taking the limit of equations (\ref{q_1N})-(\ref{q_3N}) when $\bx_{\sA}$ recedes towards the source of the light ray at infinity, i.e. when $r_{\sA}\rightarrow\infty$, $\bn_{\sA}\rightarrow -\bN_{e}$ and $\bN_{\sAB}\rightarrow\bN_{e}$. Taking into account that $\arccos \bn_{\sA}.\bn_{\sB}\rightarrow \pi-\arccos \bN_{e}.\bn_{\sB}$ when $\bn_{\sA}\rightarrow -\bN_{e}$, we get the system of equations (\ref{q1inf})-(\ref{q3inf}). QED.

The expression found for $q_3$ is new, whereas the expressions obtained for $q_1$ and $q_2$ coincide with previous results found in \cite{teyssandier2}. 

\section{Conclusion} \label{concl}

This paper is devoted to the study of the time transfer function ${\cal T}$ and the impact parameter $b$ corresponding to a photon travelling along a quasi-Minkowskian light ray in a static, spherically symmetric spacetime. The main results are the following:

{\it a)} The system of equations (\ref{TnFn})-(\ref{FABn}) enabling, at least in principle, to determine the perturbation terms ${\cal T}^{(n)}$ involved in the expansion of the time transfer function in power series of $G$. Such a system appreciably simplifies the approach developed in \cite{teyssandier1} for static, spherically symmetric spacetimes. 

{\it b)} The demonstration of the analyticity of the functions ${\cal T}^{(n)}$ when the emission and reception points are neither coincident, nor located in diametrically opposite directions.

{\it c)} The replacement of the recurrence method outlined in proposition \ref{itTAB} by an iterative procedure for solving the integro-differential equation satisfied by the time transfer function. This procedure presents the great advantage that only elementary integrations which can be performed with any symbolic computer program are required whatever the order of approximation. The legitimacy of this approach essentially rests on the property of analyticity of the functions ${\cal T}^{(n)}$. 

{\it d)} The explicit calculation of the time transfer function and the impact parameter up to the third-order terms in $G$.  The new results brought by equations (\ref{T3}) and (\ref{q_3N}) illustrate the efficiency of the procedure. The expressions obtained for the impact parameter up to the third order are extended to the case of a light ray emitted at infinity in a given direction. The ability to recover the expressions of  ${\cal T}^{(1)}$ and ${\cal T}^{(2)}$ found in previous works confirms the reliability of the method developed in this paper. 

{\it e)} A new derivation of the enhanced terms up to the order $G^3$, obtained from our full expressions of ${\cal T}^{(1)}$, ${\cal T}^{(2)}$ and ${\cal T}^{(3)}$. It is shown that the third-order enhanced term must be taken into account for determining $\gamma$ at a level of accuracy of $10^{-8}$. Surprisingly, for light rays grazing the Sun, this term is found to be larger than the first-order Lense-Thirring effect due to the solar rotation. 

Finally, it may be noted that in accordance with equations (40) and (41) in \cite{leponcin1}, our formulae would allow to derive the triples characterizing the direction of a light ray at its points of emission and reception up to the third order; these triples would enable to determine the frequency shifts. Furthermore, the calculations developed here could be extended to light rays propagating in the equatorial plane of an axisymmetric, rotating body (a Kerr spacetime, e.g.). 


\section*{Acknowledgements}

We thank one of the anonymous referees for having suggested to discuss the relevance of our results for modelling solar system experiments.

\appendix

\section{Calculation of ${\cal T}^{(3)}(\bx_{\sA},\bx_{\sB})$}

To begin with, it may be noted that (\ref{R3}) can be rewritten in the form
\begin{eqnarray} 
\fl \frac{1}{r_c^3}U_3^{\ast}=\frac{\kappa_3}{r^2s} -\frac{(1+\gamma) \kappa}{s^3}+ \frac{\kappa q_1r_c}{r s^3}+\frac{(1+\gamma)[(1+\gamma)^2+q_1^2]r^2}{2s^5} \nonumber \\
-\frac{[3(1+\gamma)^2+q_1^2]q_1r_c r}{2s^5}+\frac{(1+\gamma)q_1^2 r_c^2}{s^5}. \label{R3b}
\end{eqnarray}
It is easily seen that 
\begin{eqnarray}
\fl \int_{r_{\sA}}^{r_{\sB}}\left[\frac{\kappa_3}{r^2s} -\frac{(1+\gamma) \kappa}{s^3}+ \frac{\kappa q_1r_c}{r s^3}\right]dr \nonumber \\
=\frac{1}{r_{\sA}r_{\sB}}\left( \frac{1}{r_{\sA}}+\frac{1}{r_{\sB}}\right)\frac{\vert\bx_{\sB}-\bx_{\sA}\vert}{1+\mu}\left[\kappa_3 - (1+\gamma)\kappa \frac{\arccos \mu}{\sqrt{1-\mu^2}}\right]. \label{IR3a}
\end{eqnarray}
The calculation of the three other integrals is more lengthy. For the sake of brevity, it is convenient to put
\be \label{Rp}
R=\vert\bx_{\sB}-\bx_{\sA}\vert , \quad p=r_{\sA}r_{\sB}.
\ee
Taking into account (\ref{Nn1A}) and (\ref{Nn1B}), and then using relations as follow  
\begin{eqnarray}
\fl (r_{\sB} -r_{\sA}\mu)^3-(r_{\sB}\mu-r_{\sA})^3=(1-\mu)(r_{\sA}+r_{\sB})[R^2(1+\mu+\mu^2)\nonumber \\
-p(1+2\mu-\mu^2-2\mu^3)],\label{rr3}\\
\fl r_{\sB}^3(r_{\sB} -r_{\sA}\mu)^3-r_{\sA}^3(r_{\sB}\mu-r_{\sA})^3=R^2[R^4+3pR^2\mu-3p^2(1-\mu^2)], \label{rBrA3} \\
\fl q_1^2=(1+\gamma)^2\frac{R^2+2p(1+\mu)}{R^2}\,\frac{1-\mu}{1+\mu},
\end{eqnarray}
we are led to
\begin{eqnarray} 
\fl \int_{r_{\sA}}^{r_{\sB}}\left[\frac{(1+\gamma)[(1+\gamma)^2+q_1^2]r^2}{2s^5} 
-\frac{[3(1+\gamma)^2+q_1^2]q_1r_c r}{2s^5}+\frac{(1+\gamma)q_1^2 r_c^2}{s^5}\right]dr \nonumber \\
=\frac{1}{3r_{\sA}r_{\sB}}\left( \frac{1}{r_{\sA}}+\frac{1}{r_{\sB}}\right)\frac{\vert\bx_{\sB}-\bx_{\sA}\vert}{(1+\mu)^2} \frac{(1+\gamma)^3 I}{[(r_{\sB} -r_{\sA}\mu)(r_{\sB}\mu-r_{\sA})]^3}, \label{IR3b}
\end{eqnarray}
where $I$ is given by the lengthy expression 
\begin{eqnarray}
\fl I=R^2[R^2+p (1-\mu^2)][R^2(1+\mu+\mu^2)-p(1+2\mu-\mu^2-2\mu^3)] \nonumber \\
-(1-\mu^2)[R^2(2+\mu)+p(1-\mu^2)][R^4+3pR^2\mu-3p^2(1-\mu^2)]\nonumber\\
+(1-\mu)^2[R^2+2p(1+\mu)]\nonumber \\
\times [R^4(1+2\mu)-pR^2(1-3\mu-4\mu^2)-3p^2(1+\mu-\mu^2-\mu^3)]. \label{I}
\end{eqnarray}
Taking into account the relation
\be \label{rr}
(r_{\sB} -r_{\sA}\mu)(r_{\sB}\mu-r_{\sA})=R^2 \mu-p (1-\mu^2), 
\ee
it is easily checked that 
\be \label{I2}
\frac{I}{[(r_{\sB} -r_{\sA}\mu)(r_{\sB}\mu-r_{\sA})]^3}=3,
\ee
a result which proves to be spectacularly simple in spite of the apparent complexity of the right-hand side of (\ref{I}). Equations (\ref{IR3a}), (\ref{IR3b}) and (\ref{I2}) yield (\ref{T3}).



\end{document}